\documentclass{ws-ijmpcs}

\newcommand{\be}{\begin{equation}}
\newcommand{\ee}{\end{equation}}
\newcommand{\beann}{\begin{eqnarray*}}
\newcommand{\eeann}{\end{eqnarray*}}
\newcommand{\bea}{\begin{eqnarray}}
\newcommand{\eea}{\end{eqnarray}}

\newcommand{\bdm}{\begin{displaymath}}
\newcommand{\edm}{\end{displaymath}}

\usepackage{epsfig}

\begin{document}

\markboth{Fabris {\it et al}}
{Rastall cosmology and the $\Lambda$CDM model}

%
\catchline{}{}{}{}{}
%

\title{Rastall cosmology\footnote{Based on a talk presented by J\'ulio C. Fabris during the IWARA 2011 conference.}
}

\author{J\'ulio C. Fabris$^{b}$, Oliver F. Piattella$^{b}$, Davi C. Rodrigues$^{b}$, Carlos E. M. Batista$^a$ and Mahamadou H. Daouda$^b$}

\address{$^a$ Departamento de F\'{\i}sica, Universidade Estadual de Feira de Santana,\\
Feira de Santana, Brazil}
\address{$^b$ Departamento de F\'{\i}sica, Universidade Federal do Esp\'{\i}rito Santo,\\
Vit\'oria, Brazil}

\maketitle


\begin{history}
\received{Day Month Year}
\revised{Day Month Year}
\end{history}

\begin{abstract}
We review the difficulties of the generalized Chaplygin gas model to fit observational data, due to the tension between background and perturbative tests. We argue that such issues may be circumvented by means of a self-interacting scalar field representation of the model. However, this proposal seems to be successful only if the self-interacting scalar field has a non-canonical form. The latter can be implemented in Rastall's theory of gravity, which is based on a modification of the usual matter conservation law. We show that, besides its application to the generalized Chaplygin gas model, other cosmological models based on Rastall's theory have many interesting and unexpected new features.
\end{abstract}

\ccode{PACS numbers: 04.50.Kd, 95.35.+d, 95.36.+x, 98.80.-k}

\vspace{.4in}

Large scales observations indicate that about $95\%$ of the matter content is composed of two exotic components, one (dark matter) with zero effective pressure and responsible for the formation of local structures, and the other (dark energy) with negative pressure and responsible for the present stage of accelerated expansion. They are exotic in the sense that no elementary particle of the standard model is able to represent them. Indeed, they do not emit any kind of electromagnetic radiation and up to now were only indirectly detected through their gravitational effects.

The conventional cosmological model contains the usual known components (baryons, radiation and neutrinos, understood as all the particles forming the standard model), supplemented by dark matter and dark energy. The former is represented by a pressureless component that decouple from radiation much earlier than the baryons,\cite{bertone} whereas dark energy is modelled by a cosmological constant $\Lambda$, allegedly representing vacuum energy associated to quantum fields.\cite{caldwell} The standard model of cosmology is named $\Lambda$CDM.

The most important sign of the success of the $\Lambda$CDM is its ability to nicely fit both kinematic (i.e. regarding the background evolution history) and dynamical (i.e. regarding perturbations about a homogeneous and isotropic background) cosmological observations.
The free parameters are essentially three: $H_0$, i.e. the value of the Hubble parameter today, $\Omega_{\rm m0}$ and $\Omega_{\Lambda 0}$, i.e. the dark matter and dark energy densities parameters. Under the condition that the universe is spatially flat, i.e $\Omega_{\rm m0} + \Omega_{\Lambda 0} = 1$, the number of free parameters is reduced to two. Recent analysis of CMB observations,\cite{komatsu} obtain $H_0 = 70.4^{+1.3}_{-1.4}$ km s$^{-1}$ Mpc$^{-1}$, $\Omega_{\Lambda0} = 0.728^{+0.015}_{-0.016}$ and $\Omega_{\rm m0} =0.227 \pm 0.014$. Moreover, the baryon's fraction to the critical density is given by $\Omega_{\rm b0} = 0.0456 \pm 0.0016$ and the curvature parameter of the spatial section is $-0.0133 < \Omega_{\rm k0} < 0.0084$, indicating an almost flat universe.

In spite of its success, the $\Lambda$CDM model faces many difficulties, both theoretical and observational. The first one concerns the nature of the components of the dark sector. The most quoted candidates to represent dark matter are neutralinos (stable supersymmetric particles) and axions (predicted by the Grand Unified models). No one of the underlying fundamental theory for these candidates has received experimental evidence until now, thus they remain theoretical proposals. In some sense, evidences for the existence
of dark matter may be considered as an indirect proof that such fundamental theories may have something to do with nature. On the other hand, the dark energy component, the cosmological constant, is usually interpreted as vacuum energy emerging from quantum fields in curved spacetime. However, theoretical estimates lead to values up to 120 orders of magnitude larger than its observed value. On the other hand, if considered as a pure geometric term, the observed value of the cosmological constant requires an extremely precise tuning of its value.

Moreover, there are at least two other observational aspects that bring troubles to the $\Lambda$CDM model. The first one is the so-called \textit{cosmic coincidence} and concerns the fact that the present fractional densities for dark matter and dark energy are of the same order. This is very unlikely to happen since dark energy and dark matter scale very differently with time. Second, predictions of the $\Lambda$CDM model for non-linear structures in the universe do not fit numerical simulations, mainly in what concerns the existence of an excess of power in the mass agglomeration at galactic scales.\cite{Primack:2009vp}

In view of these problems, many alternatives to the $\Lambda$CDM model have been proposed.\cite{padma} For example, quintessence and $K$-essence. However, both have their specific problems, like the extreme small value of the scalar particle in the quintessence model (i.e. $m_\phi \sim 10^{-33}$ eV) and the stability issues for the $K$-essence model. An interesting proposal is the interacting models for the dark sector, which consider the possibility of the decay of one component into the other, and may shed some light on the cosmic coincidence problem. On the other hand, they suffer from the necessity of introducing phenomenological interaction terms, due to the absence of a fundamental model for the decaying process. At the same time, thermodynamical aspects concerning such interaction processes lead to constraints that seem to be in contradiction with observations.\cite{pavon,chineses}
 
Among the alternatives evoked to describe the dark sector of the universe, one interesting proposal are the so-called unified models. Their prototype is the Chaplygin gas (CG),\cite{moschella} which is inspired by string theory: it is a fluid able to mimic dark matter in the past and dark energy in the future, thus properly reproducing the expansion history of the universe. However, the CG suffers from some difficulties when it is tested against observation. This led to a more general formulation, the so-called generalized Chaplygin gas (GCG),\cite{berto1,neven} which possesses a new parameter $\alpha$ ($\alpha = 1$ recovers the original CG). Unfortunately, a tension appears between the predictions of the model and various observational tests: the background ones (e.g. SNIa, BAO, H) favour $\alpha < 0$,\cite{colistete,paulo} whereas the perturbative ones require $\alpha > 0$ in order to have a positive speed of sound.\cite{zimdahl,berto2,finelli,piattella1}

In order to cope with this problem a possibility is to abandon the fluid description for the GCG, using instead a self-interacting scalar field. Unfortunately, this cannot be done by using a canonical scalar field. We show in what follows that an interesting way out is to use a non-canonical self-interacting scalar field as suggested by Rastall's theory of gravity, an alternative
to General Relativity where the imposition of a strict conservation of the energy-momentum tensor is relaxed.\cite{rastall} Besides the alternative description for the GCG model, Rastall's theory may lead to many interesting cosmological scenario, as we describe below.

Consider the GCG equation of state:
\begin{equation}
\label{eos}
p = - \frac{A}{\rho^\alpha}\;,
\end{equation}
where $\rho$ is the density and $\alpha$ and $A$ are positive parameters. Inserting this expression in the continuity equation in a Friedmann-Lema\^{\i}tre-Robertson-Walker (FLRW) metric, i.e.
\begin{equation}
\dot\rho + 3\frac{\dot a}{a}(\rho + p) = 0\;,
\end{equation}
where $a$ is the scale factor and the dot indicates derivative with respect to the cosmic time,
one finds the following expression:
\begin{equation}
\rho = \rho_0\left[\bar{A} + (1 - \bar{A})a^{-3(1 + \alpha)}\right]^\frac{1}{1 + \alpha}\;,
\end{equation}
where $\bar{A} \equiv A/\rho_0^{1 + \alpha}$ and $\rho_0$ is the present-time GCG density (the present time is fixed by requiring $a_0 = 1$).
\par
The tension between background and perturbative constraints can be easily noticed from the equation of state (\ref{eos}): while in principle such definition allows for any real value of $\alpha$, at the perturbative level we have to deal with the square of the speed of sound, which has the following form:
\begin{equation}
c_s^2 \equiv \frac{d p}{d\rho} = \frac{\alpha A}{\rho^{1 + \alpha}}\;,
\end{equation}
which is positive only for $\alpha > 0$.
\par
The source of the tension comes from the fact that background tests favour negative values of $\alpha$,\cite{colistete} which are excluded at perturbative level due to the occurrence of an imaginary speed of sound.\cite{zimdahl} This fact is due to the fluid representation adopted. It must be noted that such fluid representation emerges naturally from the Nambu-Goto action in the original CG formulation.\cite{jackiw} The DBI scalar field formulation of the Chaplygin gas is completely equivalent to the fluid formulation, even at the perturbative level.\cite{eduardo,thais} On the other hand, the GCG has not the same clear connection with string theory (even if a DBI-type action can be written presenting similar properties to the strict DBI case). In consequence, there is no fundamental reason to stay in the context of string theory. Therefore, it is possible to conceive a more fundamental representation for the GCG. A first possibility is to interpret the CGC as a self-interacting scalar field. However, a canonical self-interacting scalar field possesses a speed of sound equal to the speed of light, as we prove below, limiting the possibility to use such description for a unified model of dark matter and dark energy.
\par
The adiabatic speed of sound of a generic $K$-essence scalar field has the following expression:\cite{neven}
\begin{equation}\label{BDformula}
\hat c_s^2 = \frac{p_{,\chi}}{\rho_{,\chi}}
\end{equation}
where $\chi$ is the kinetic term.
A canonical self-interacting scalar field has the following expressions for the pressure and the density in the FLRW background:
\begin{equation}
\rho_\phi = \frac{\dot\phi^2}{2} + V(\phi)\;, \quad p_\phi = \frac{\dot\phi^2}{2} - V(\phi)\;,
\end{equation}
and it obeys the Klein-Gordon equation, which, in the FLRW background, reads
\begin{equation}
\ddot\phi + 3H\dot\phi = - V_\phi(\phi)\;,
\end{equation}
where $V_\phi(\phi) \equiv dV/d\phi$. Hence, the adiabatic speed of sound for a canonical self-interacting scalar field is $\hat{c_s}^2 = 1$. This proves that a canonical scalar field cannot play the role of dark matter.
\par
So, how a self-interacting scalar field could represent the dark matter component? We have to consider a non-canonical scalar field. We adopt Rastall's gravity theory,\cite{rastall} where the energy-momentum tensor $T^{\mu\nu}$ does not obey the usual conservation law. Instead, the divergence of $T^{\mu\nu}$ reads
\begin{equation}
{T^{\mu\nu}}_{;\mu} = \kappa R^{;\nu}\;,
\end{equation}
where $R$ is the Ricci scalar and $\kappa$ is a constant. The field equations in Rastall's theory read
\begin{eqnarray}
\label{fe}
R_{\mu\nu} - \frac{1}{2}g_{\mu\nu}R &=& 8\pi G\left(T_{\mu\nu} - \frac{\gamma - 1}{2}g_{\mu\nu}T\right)\;,\\
\label{cons-law}
{T^{\mu\nu}}_{;\mu} &=& \frac{\gamma - 1}{2}T^{;\nu}\;,
\end{eqnarray}
where $\gamma$ is a dimensionless constant connected to $\kappa$. When $\gamma = 1$, general relativity is recovered. Using the canonical form for the energy-momentum tensor of a scalar field, i.e.
\begin{equation}
T_{\mu\nu} = \phi_{,\mu}\phi_{,\nu} - \frac{1}{2}g_{\mu\nu}\phi_{,\rho}\phi^{,\rho} + g_{\mu\nu}V(\phi)\;,
\end{equation}
we obtain the following coupled equations:
\begin{eqnarray}
\label{Ein00} R_{\mu\nu} - \frac{1}{2}g_{\mu\nu}R &=& \phi_{,\mu}\phi_{,\nu} - \frac{2 - \gamma}{2}g_{\mu\nu}\phi_{,\alpha}\phi^{,\alpha} + g_{\mu\nu}(3 - 2\gamma)V(\phi)\;,\\
\label{Einss}
\Box\phi + (3 - 2\gamma)V_{,\phi} &=& (1 - \gamma)\frac{\phi^{,\rho}\phi^{,\sigma}\phi_{;\rho;\sigma}}{\phi_{,\alpha}\phi^{,\alpha}}\;.
\end{eqnarray}
From Eq.~\eqref{Ein00}, the following effective energy-momentum tensor can be read off:
\begin{equation}
\label{efetivo}
T_{\mu\nu}^{\rm eff} = \phi_{,\mu}\phi_{,\nu} - \frac{2 - \gamma}{2}g_{\mu\nu}\phi_{,\alpha}\phi^{,\alpha} + g_{\mu\nu}(3 - 2\gamma)V(\phi)\;,
\end{equation}
implying the following expressions for the energy density and pressure:
\begin{equation}
\rho_\phi^{\rm eff} = \frac{\gamma}{2}\dot\phi^2 + (3 - 2\gamma)V(\phi)\quad, \quad p_\phi^{\rm eff} = \frac{2 - \gamma}{2}\dot\phi^2 - (3 - 2\gamma)V(\phi)\;.
\end{equation}
Using this expression in Eq.~\eqref{BDformula} to evaluate the speed of sound, one finds
\begin{equation}
c_s^2 = \frac{2 - \gamma}{\gamma}\;.
\end{equation}
This implies a vanishing speed of sound for $\gamma = 2$. In this case, the non-canonical self-interacting scalar field based on Rastall's theory may represent dark matter. On the other side, from the non-perturbative point of view, it can represent dark energy by a suitable choice of the potential. We will explore this possibility in what follows.
\par
Let us consider a self-interacting scalar field, with the effective energy-momentum tensor (\ref{efetivo}), with $\gamma = 2$:
\begin{equation}
T_{\mu\nu}^{\rm eff} = \phi_{,\mu}\phi_{,\nu} -  g_{\mu\nu}V(\phi)\;.
\end{equation}
Inserting the FLRW metric, we have the following density and pressure associated with this scalar field:
\begin{equation}
\rho_\phi = \dot\phi^2 - V(\phi)\;, \quad p_\phi = - V(\phi)\;.
\end{equation}
Let us suppose that this density and pressure reproduce the background behaviour of the GCG model. Hence, in this case, we have,
\begin{eqnarray}
\dot\phi(a) &=& \sqrt{3\Omega_{c0}}\sqrt{g(a)^{1/(1 + \alpha)} - \bar A g(a)^{-\alpha/(1 + \alpha)}}\;,\\
V(a) &=& 3\Omega_{c0}\bar A g(a)^{-\alpha/(1 + \alpha)}\;,
\end{eqnarray}
where $g(a) \equiv \bar A + (1 - \bar A)a^{-3(1 + \alpha)}$. Hence, in order to have a zero speed of sound, the scalar model must obey the following equations:
\begin{eqnarray}
 R_{\mu\nu} &-& \frac{1}{2}g_{\mu\nu}R = 8\pi GT_{\mu\nu} + \phi_{,\mu}\phi_{,\nu} + g_{\mu\nu}V(\phi)\;,\\
 \Box\phi &+& V_\phi + \frac{\phi^{,\rho}\phi^{,\sigma}\phi_{;\rho;\sigma}}{\phi_{,\alpha}\phi^{,\alpha}} = 0\;,
\end{eqnarray}
where, just for future convenience, we have made the redefinition $V(\phi) \rightarrow - V(\phi)$. 
\par
Let us inspect now the perturbative behaviour of this system, computing scalar perturbations in the density contrast.
The perturbed equations in the synchronous coordinate condition read:\cite{thais}
\begin{eqnarray}
\label{rastall1}
\ddot\delta &+& 2\frac{\dot a}{a}\dot\delta - \frac{3}{2}\frac{\Omega_0}{a^3}\delta =
\dot\phi\dot\Psi - V_\phi\Psi\;,\\
\label{rastall2}
2\ddot\Psi &+& 3\frac{\dot a}{a}\dot\Psi + \left(\frac{k^2}{a^2} + V_{\phi\phi}\right)\Psi = \dot\phi\dot\delta\;,
\end{eqnarray}
where $\Psi = \delta\phi$ and $\delta$ is the density contrast of the matter component. Using the scale factor as independent variable, the above system of equations takes on the following form:
\begin{eqnarray}
 \delta'' &+& \left[\frac{2}{a} + \frac{f'(a)}{f(a)}\right] \dot\delta - \frac{3}{2}\frac{\Omega_0}{a^3f^2(a)}\delta =
\phi'\Psi' - \frac{V_\phi}{f^2(a)}\Psi\;,\\
2\Psi'' &+& \left[\frac{3}{a} + 2\frac{f'(a)}{f(a)}\right]\Psi' + \left[\frac{k^2}{a^2f^2(a)} + \frac{V_{\phi\phi}}{f^2(a)}\right]\Psi = \phi'\delta'\;,
\end{eqnarray}
where $f(a) = \dot a = \sqrt{\Omega_{m0}a^{-1} + \Omega_c(a)a^2}$ and $\Omega_c(a) = \Omega_{c0}g(a)^{1/(1 + \alpha)}$.
\par
Using a Bayesian analysis and comparing the theoretical predictions of our model with the 2dFRGS data for the power spectrum of matter distribution in the universe, we find a significant probability region 
for $\alpha < 0$. The results are shown in figure 1, for the unification scenario where the dark sector is represented by the GCG only.\cite{thais}

\begin{figure}
\begin{center}
\includegraphics[width=0.45\linewidth]{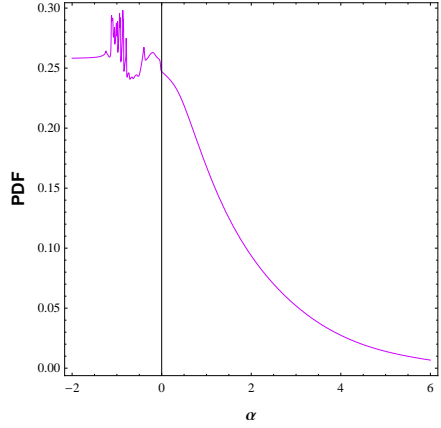}
\label{rastall-fig2}
\end{center}
\caption{PDF for the GCG parameter $\alpha$ for the unification scenario, i.e., $\Omega_{\rm dm0} = 0$.} 
\end{figure}

Hence, if the GCG model is represented by a non-canonical scalar field, like the one suggested by Rastall's theory of gravity, 
the observational tension that plagues the GCG fluid model may disappear or, at least, be considerably alleviated. This fact may open new perspectives for the dark matter-dark energy unification program.

However, this non-canonical scalar formulation obeying the Rastall's conservation equation has some striking peculiarities. Let us consider, for example, scalar perturbations in the Newtonian gauge:
\begin{eqnarray}
ds^2 = a^2(1 - 2\Phi)d\eta^2 - a^2(1 + 2\Phi)\gamma_{ij}dx^idx^j, \nonumber
\end{eqnarray}
The perturbed Rastall's equations (\ref{Ein00})-(\ref{Einss}) becomes:
\begin{eqnarray}
 \nabla^2\Phi - 3\mathcal{H}\left(\mathcal{H}\Phi + \Phi'\right) + \gamma\left(\mathcal{H}^2 - \mathcal{H}'\right)\Phi =\nonumber\\ 4\pi G\left[\gamma\phi_0'\delta\phi' + (3 - 2\gamma)a^2V_{,\phi}\delta\phi\right]\;,\\
\label{G0i} \mathcal{H}\Phi_{,i} + \Phi'_{,i} = 4\pi G\phi_0'\delta\phi_{,i}\;,\\
\Phi'' + 3\mathcal{H}\Phi' + \left(2\mathcal{H}' + \mathcal{H}^2\right)\Phi + (2 - \gamma)\left(\mathcal{H}^2 - \mathcal{H}'\right)\Phi =\nonumber\\ 4\pi G\left[(2 - \gamma)\phi_0'\delta\phi' - (3 - 2\gamma)a^2V_{,\phi}\delta\phi\right]\;.
\end{eqnarray}
These equations can be combined in different ways in order to obtain a single equation for the potential $\Phi$. These different combinations lead
to the following equations:
\begin{eqnarray}
\label{Eq1} \Phi'' + 3\mathcal{H}\Phi' + \left(2\mathcal{H}' + \mathcal{H}^2\right)\Phi =\nonumber\\ \frac{2 - \gamma}{\gamma}\left[-k^2\Phi - 3\mathcal{H}\left(\mathcal{H}\Phi + \Phi'\right)\right] - \frac{2V_{,\phi}a^2}{\gamma\phi_0'}(3 - 2\gamma)\left(\mathcal{H}\Phi + \Phi'\right)\;,
\\
 \Phi'' + 3\mathcal{H}\Phi' + \left(2\mathcal{H}' + \mathcal{H}^2\right)\Phi =\nonumber\\ -k^2\Phi - 3\mathcal{H}\frac{2 - \gamma}{\gamma}\left(\mathcal{H}\Phi + \Phi'\right) - \frac{2V_{,\phi}a^2}{\gamma\phi_0'}(3 - 2\gamma)\left(\mathcal{H}\Phi + \Phi'\right)\;.\label{Eq2}
\end{eqnarray}
These equations are identical only in two cases: either $\gamma = 1$ or if the perturbed quantities are not space dependent, leading
to $k = 0$. In the first case, the Rastall's theory reduces to general relativity. In the second case we have a pure redefinition of the background quantities, remaining with a homogeneous universe.\cite{plb}
\par
A way to give sense to perturbations in the scalar formulation of the Rastall's theory without reducing it to general relativity, is to consider, besides the scalar field, a hydrodynamic fluid. In such situation, the problem presented above disappear
and perturbations can be consistently defined for any value of $\gamma$. Hence, this non-canonical scalar field requires matter in order to make sense perturbatively.

This is not the only surprise that results from the new conservation law (\ref{cons-law}). In some situations, the Rastall's cosmology can lead to the same achievements of the $\Lambda$CDM model, but with new features at the non-linear regime where the $\Lambda$CDM model faces some difficulties.\cite{prd}

Let us consider a two fluid model, with one of the components obeying the traditional conservation law. In this case, equation (\ref{cons-law})
splits into two equations:
\begin{eqnarray}
{T^{\mu\nu}_x}_{;\mu} = \frac{\gamma - 1}{2}T^{;\nu}\;, \qquad {T^{\mu\nu}_m}_{;\mu} = 0\;,
\end{eqnarray}
where the subscripts $m$ and $x$ indicate the matter and dark energy components. Choosing $p_m = 0$ et $p_x = - \rho_x$, we find the following equations, considering also equation (\ref{fe}):

\begin{eqnarray}
H^2 &=& \frac{8\pi G}{3}\left[(3 - 2\gamma)\rho_x + \frac{-\gamma + 3}{2}\rho_m\right]\;, \\
\dot\rho_m + 3H\rho_m &=& 0\;, \qquad (3 - 2\gamma)\dot\rho_x = \frac{\gamma - 1}{2}\dot\rho_m\;.
\end{eqnarray}
Integrating for the densities, we obtain,
\begin{eqnarray}
\rho_m = \frac{\rho_{m0}}{a^3}\;,\qquad \rho_x = \frac{\rho_{x0}}{3 - 2\gamma} + \frac{\gamma - 1}{2(3 - 2\gamma)}\rho_m\;.
\end{eqnarray}
Inserting these expressions in the modified Friedmann equation, it results
\begin{eqnarray}
H^2 &=& \frac{8\pi G}{3}(\rho_{x0} + \rho_m)\;,
\end{eqnarray}
that is, Friedmann equation for the $\Lambda$CDM model. In spite of this, the dark energy component is now dynamical and scales with
time as
\begin{eqnarray}
\rho_x = \frac{\rho_{x0}}{3 - 2\gamma} + \frac{\gamma - 1}{2(3 - 2\gamma)}\rho_m\;.
\end{eqnarray}
This means that all the successes of the $\Lambda$CDM model in fitting the so-called kinematic observation tests (Supernova type Ia, BAO, $H(z)$), are recovered by such Rastall's cosmology.

Astonishingly, the $\Lambda$CDM structure with a dynamical dark energy component is preserved at perturbative level. In fact, considering linear perturbations (for simplicity, we consider now the synchronous gauge condition), after standard calculations, we end up with a single equation for the matter density contrast $\delta_m = \delta\rho_m/\rho_m$:
\begin{eqnarray}
\ddot\delta_m + 2\frac{\dot a}{a}\delta_m - 4\pi G\rho_m\delta_m = 0\;.
\end{eqnarray}
It is again the same equation of the $\Lambda$CDM model. Hence, in what concerns the dynamical cosmological tests, requiring perturbative analysis (like the CMB anisotropy and matter power spectrum), Rastall's theory is perfectly equivalent to the $\Lambda$CDM model.

But, there exist fluctuations in the dark energy component given by,
\begin{eqnarray}
\delta\rho_x = \frac{\gamma - 1}{2(3 - 2\gamma)}\delta\rho_m\;.
\end{eqnarray}
In order words, now dark energy agglomerates. Such structure is preserved even at second order perturbations, when neglecting vector and tensor contributions.\cite{prd}

This agglomeration of the dark energy component should have important consequences at non-linear level. For example, let us consider the spherical collapse model. The equation describing the evolution of a spherical perturbation is given by,
\begin{eqnarray}
\biggr(\frac{\dot a}{a_i}\biggl)^2 = H_i^2\left[\Omega_p(t_i)\frac{a_i}{a} + 1 - \Omega_p(t_i)\right]\;,
\end{eqnarray}
where $a_i$ is the scale factor computed at some initial time $t_i$ at which the collapse begins and $\Omega_p = 1 + \delta$ is the
density parameter of the collapsing region, determined by the density contrast of the fluid there contained. In the Rastall's cosmology described above, we have
\begin{eqnarray}
\Omega_p = 1 + \frac{2(3 - 2\gamma)}{5 - 3\gamma}\delta_m + \frac{\gamma - 1}{5 - 3\gamma}\delta_x\;,
\end{eqnarray}
where the perturbations on the dark energy component appear. This term is, of course, absent in the corresponding expression for
the $\Lambda$CDM model, as can be seen by setting $\gamma = 1$ in the relation above.

All the features described here involving the Rastall's cosmology, reveal the richness of such a proposal based on the modification of the usual conservation law. However, it is clear that a deeper analysis is required, possibly in connection with quantum effects in the universe. At the same time, the implications of Rastall's cosmology for the formation and properties of non-linear structures is a very promising research program.

\section*{Acknowledgements}

We thank CNPq (Brazil) for partial financial support. J.C.F. thanks the organizers of the conference IWARA2011
for their kind hospitality.

\end{document}